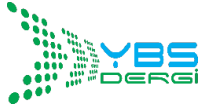

**YÖNETİM BİLİŞİM SİSTEMLERİ DERGİSİ**
http://dergipark.gov.tr/ybs

# VIDEO OR IMAGE TRANSMISSION SECURITY FOR ESP-EYE IoT DEVICE USED IN BUSINESS PROCESSES

Ömer AYDIN[1*], İbrahim İsmail ERHAN[2]

[1] Assistant Professor Doctor, Faculty of Engineering, Manisa Celal Bayar University, Yunusemre, Manisa, Turkey
omer.aydin@cbu.edu.tr ORCID: 0000-0002-7137-4881
[2] Computer Engineering Department, Dokuz Eylül University, İzmir, Turkey, ibrahim.erhan@ceng.deu.edu.tr
ORCID: 0000-0001-6290-9437
[*]**Corresponding author:** omer.aydin@cbu.edu.tr

**ABSTRACT**

Internet of Things is the name of a communication network that is formed by physical objects such as RFID tags, sensors and some lightweight development platforms that have the ability to connect to the internet. While the devices can communicate among themselves in this network, they can also be part of a large network. The data produced by those physical objects which are the member of IoT network are processed by different methods and the outputs obtained are used in processes such as decision making and learning. With this aspect of the Internet of Things, it affects all areas of human life and its number is increasing day by day. These devices appear to have security gaps due to their limited resources, their wide range of usage area and incomplete security standards. These devices, which are located in people's living areas, manufacturing and business processes also cause difficulties in protecting privacy. In this study, a solution has been developed for the communication security of the internet of things called ESP-Eye which includes a camera, wireless communication module and face recognition software. The proposed solution was implemented on the ESP-Eye.

# İŞ SÜREÇLERİNDE KULLANILAN ESP-EYE IoT CİHAZI İÇİN VİDEO VEYA GÖRÜNTÜ İLETİM GÜVENLİĞİ

**ÖZET**

Nesnelerin İnterneti, RFID etiketleri, sensörler, internete bağlanma özelliğine sahip bazı hafif geliştirme platformları gibi fiziksel nesnelerin oluşturduğu bir iletişim ağının adıdır. Bu ağda cihazlar kendi aralarında haberleşebilirken, aynı zamanda büyük bir ağın parçası olabilirler. Nesnelerin interneti ağının üyesi olan bu fiziksel nesnelerin ürettiği veriler farklı yöntemlerle işlenir ve elde edilen çıktılar karar verme ve öğrenme gibi süreçlerde kullanılır. Nesnelerin İnternetinin bu yönü ile insan hayatının her alanını etkilemekte ve sayısı her geçen gün artmaktadır. Bu cihazlar, sınırlı kaynakları, geniş kullanım alanları ve eksik güvenlik standartları nedeniyle güvenlik açıklarına sahiptirler. İnsanların yaşam alanlarında, imalat ve iş süreçlerinde kullanılan bu cihazlar, mahremiyetin korunmasında da zorluklara neden oluyor. Bu çalışmada ESP-Eye adı verilen Nesnelerin İnterneti cihazında haberleşme güvenliği için kamera, kablosuz iletişim modülü ve yüz tanıma yazılımı içeren bir çözüm geliştirilmiştir. Önerilen çözüm ESP-Eye'da uygulanmıştır.
**Anahtar Kelimeler:** ESP-Eye, Görüntü, Akış, Güvenlik, İş süreçleri

**INTROCUTION**

Internet of Things (IoT) takes measurements and collects data at every point from logistics processes, assembly lines and various points of daily life. The information obtained from these devices shapes large and complex business processes (Friedow et al., 2018). A lot of research has been done on these devices in recent years. Devices such as sensors, RFID tags and actuators connect to the internet and thus become part of many internet-based services and applications. The studies conducted generally focused on the actual applications. There has not been much study done on issues that will affect business processes such as the integration of these devices with Enterprise Resource Planning systems (Meyer et al., 2013).

When the business processes are evaluated, it is seen that they are affected by external events. This also occurs especially in Internet of Things (IoT) scenarios. One of the parts of the Business Process Model and Representation (BPMN) standard is modelling constructs for different type of events. Process-oriented information systems should be supported in the integration of external events into business processes, so that the integration between the conceptual process model and the application can be achieved. Mandal et al. used a scenario in IoT areas and revealed the requirements for the integration. They proposed a framework so extended the BPMN process model to specify the type of expected events and also implemented a system that presents the integration (Mandal et al., 2017).

The use of information systems, IoT and many other new technologies in business processes has become widespread. IoT devices, on the other hand, started to appear in different sizes, forms and abilities. One of such IoT devices is ESP-Eye with built-in camera feature.

After the ease of use of IoT devices and their engagement in daily life, people started to use these devices in every environment we could imagine. For example, health, industry, homes, offices. The information collected with these devices has become a support for other systems (Giusto et al ,2010). It can be used in situations such as feeding a facial identification machine learning model or making the production in a factory more efficient. The collected data may include very important production analyzes for a company, as well as data that may create privacy if used in a normal person's home. IoT is easy to integrate with other systems, small in size and inexpensive, and can be developed by anyone with or without expertise in this field. When designing all these IoT devices, due to limited resources and processing power, security has not been considered and has become a major target in the face of threats (Seralathan et al., 2010; Apthorpe et. al., 2017). DDoS attacks are known to use IoT devices by attackers as part of botnet structures. The ease of control and the disregard of their security for the current situation made IoT devices available to attackers as booster for DDoS attacks (Kolias et.al, 2017). In addition to these, different attacks can be tried. For example, some protocols such as Network Time Protocol (NTP) and Simple Network Time Protocol (SNTP) have been tried on some IoT devices (Çepik et.al, 2020; 2021). In addition, some lightweight authentication protocols have been proposed for the security of IoT devices communication (Aydin, 2020; Aydin et.al, 2020).

IoT devices provide great opportunities in technology and create a certain infrastructure that allows people to develop projects on this infrastructure. In addition, it has brought great security problems to us before people realize it (Duncan et al., 2016). This study was presented in order to increase our awareness and not to cause this great disaster. In fact, disasters can be prevented if necessary precautions are taken.

IoT devices that are connected to the Internet and do not have any threat protection can broadcast video thanks to the cameras on it. These devices, which are open to attacks, can easily be found by attackers by scanning the whole internet. The deprivation and private information may become available to unauthorized persons. Lei, Fu, Juchem and Hogrefe stated that security problems are increasingly important and there are still devices that are weak in issues such as authentication and authorization, access control and key management for data confidentiality (Lei et al.,2005).

In this study, a reliable video broadcast system was proposed using ESP-Eye, an IoT device. The IoT device, which has a camera connected wirelessly to the network, broadcasts the recorded images with a server running on it. In this study, firstly analysis and design processes were carried out as a methodology. Then, test criteria and methods were determined. Data were collected to show the reliability of the method and then the data were analyzed. We determined our needs during the analysis phase and the basic issues such as which hardware, software and programming languages to use. We designed the system in parallel with the determination of the needs. During this period, new needs were added to the list. After meeting these needs, we implemented the system. Then we determined the necessary criteria and methods for testing the system. In particular, criteria were determined that could test the stages such as transmitting the image, encrypting and sending the encrypted text over the network. Later, the system was started to collect data. Finally, the obtained data were analyzed.

**RELATED WORKS**

Performing security operations on an IoT device differs from other systems due to certain limitations. Hamoudy, Qutqut and Almasalha stated that on the 5 main security needs IoT devices: key management, authentication, watermarking, encryption. Excessive use of processor power increases power consumption, which paves the way for a review of the methods used in encryption algorithms or the development of new encryption methods (Hamoudy et al, 2017). For example, the partial encryption method for restricted power environments is shown (Almasalha et al., 2014).

Malić, Dobrilović and Petrov proposed a new solution based on the security cameras in use, which would eliminate their functional deficiencies and reduce costs (Malić et al., 2016). Arduino and USB have installed the system using a web camera. Captured images are stored in the internal memory of the device and in the cloud. MongoDB based session management, managing user accounts, complex aggregation queries modules were used to access these images. Although the proposed system has advantages over the existing system, the security provided by the databases in terms of security and data protection alone will not be sufficient and new problems will arise. The work has already been defined as future work.

Due to the resource constraints of IoT devices, Sammoud, Kumar, Bayoumi and Elarabi mentioned the biggest difficulties of video transfer and compression operations and stated that they have security difficulties and presented a method to solve this (Sammoud et al., 2017). This method of encryption algorithms, encoding and compression operations, realizing the IoT device saves resources. In a Wee and Apostolopoulos study, a similar method was divided into two as Application-level encryption and network-level encryption (Wee and Apostolopoulos, 2001). In the application-level encryption method, the video is first encoded and compressed, and then compressed, which is divided into packets for transport in the bit stream, and in this method, there are disadvantages in case one of the packets becomes corrupted or lost during migration. In the network-level encryption method, the encoded and compressed video is separated into packets and the encryption operations are applied to these parts separately.

The ESP32-CAM module, which is an IoT device, is used for capturing and transmitting images in the system on which we will operate our security model. In similar video broadcasting systems, the ESP32 module is widely used. Raghavendra et al. used the ESP32 M5 module and the OV2640 optical camera, it installed an IoT-based surveillance system (Raghavendra et al., 2019). Ringe et al. designed smart refrigerators using Rasberry Pi and ESP32-CAM and took images from the refrigerator through the ESP32 server (Ringe et al, 2019). In this project, there are no security measures for confidentiality or transmission of sensitive images.

Kakoulin et al. used a smart camera in two modes, standby mode and recording mode (Kokoulin et al., 2019). In standby mode, it only takes images that are low enough to detect human shape. Once a human is detected, it starts broadcasting in high-resolution video. ESP32-CAM was used to perform these processes. Being cheap, having a server running on it, reducing the load on other system parts. It allows running on complex security algorithms.

**IoT USED BUSINESS PROCESSES**

We are now producing data almost everywhere in our lives. Smart watches, mobile phones, sensors and many other physical objects collect this data. For example, the smartphones become devices that can collect many information such as location, shopping habits, tastes, and holiday preferences. In parallel with the use of technology in daily life, there are similar processes in the business world. IoT devices are used in many areas such as process tracking, decision support systems, productivity enhancement, quality control and management of logistics processes. Under this section, we will examine some applications used in business processes.

Quality control processes are of great importance in businesses. Poor product quality or disruptions in the manufacturing process can have far-reaching financial consequences for any company. To minimize these risks, thermography can be used to perform quality control at appropriate businesses by purchasing a thermal camera. With this kind of camera, non-contact detection of thermal irregularities and processes can be optimized (Testo, 2021). With this application, thermal cameras can be used in quality control processes in suitable businesses.

High product and material costs, fluctuating market demand, the need for high quality and low rejection rate require new methods. The use of IoT in production is at the top of this new method. In this way, a step is taken for digital transformation in quality management. Effective quality management requires real-time controlling and monitoring of the processes, devices and people in manufacturing (BehrTech, 2021). Digital twins, supply chain management, self-dependent systems, workshop mirroring and smart pumping are some systems using IoT devices (The Manufacturer, 2018). IoT also used in retailing processes such as Customer satisfaction monitoring, food safety monitoring and asset tracking (Rawlings, 2019).

One of the secure, new, cost-effective and effective methods to control transactions in real-time and collect data is the use of cameras. With the camera systems, the processes can be visualized and continuously monitored. The use of cameras helps to eliminate the weaknesses of many real-time systems such as unreliability, unavailability and bad profitability. Previous studies on this subject have results that support this situation (Alias et al. 2015; Alias et al. 2016; Özgür et al. 2015).

**MATERIALS**

In this study, which is designed to be used in real systems or to create awareness, attention has been paid to the choice of materials used. In order to provide a complete security solution, materials which are used in the literature or have high potential for future use have been selected. In this section these materials will be given. ESP32-Cam, Arduino IDE, FDTI Programmer and ESP32 Camera Server will be investigated in subtitles below.

**ESP-EYE and ESP32-CAM**

ESP-EYE is a small IoT device with ESP32 Board, 2 MP camera and a Bluetooth module manufactured by Espressif Systems. It also features a microSD card slot that can be useful to store images taken with the camera or to store anomaly images for future analysis. This module will be the video streaming device for this study.

The ESP32-CAM is an IoT device that consumes very little power and can work with deep sleep current. It is widely used in many areas, especially in wireless control applications. IoT offers a very good solution for applications.

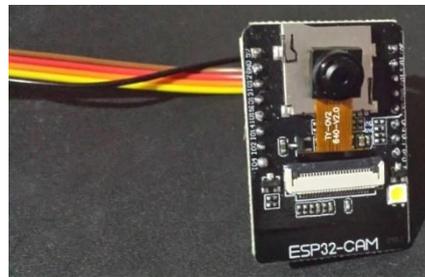

Figure 1. ESP32-CAM Board with OV2640 Camera.

We can list the main features of ESP32-CAM board shown in Figure 1;

- The smallest 802.11b/g/n Wi-Fi BT SoC module
- Low power 32-bit CPU, can also serve the application processor
- Support OV2640 camera, built-in flash lamp

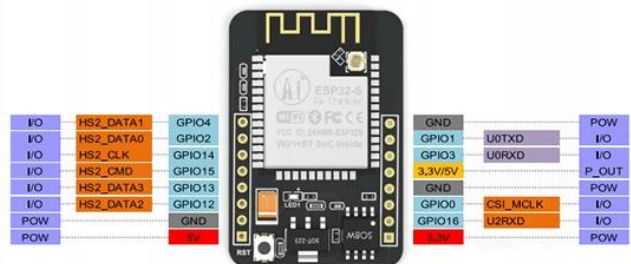

Figure 2. ESP32-CAM Pinout and Scales (ESP32-CAM Module, 2017)

Figure 2 shows ESP32-CAM's pin details. GPIO 1 and GPIO 3 pins are used for upload code data to the ESP32. GND and 3.3V pins are used for power.

**Arduino IDE**

Arduino IDE is used for code implementations in this study so ESP32 library needs to be installed into Arduino IDE to achieve that.

ESP32 board library includes camera operator functions and other essentials. ESP32 library installation screenshot can be seen in Figure 3.

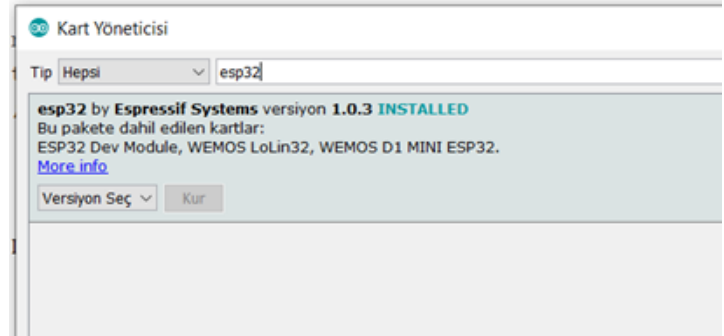

**Figure 3.** Installing the ESP32 Library

**FDTI Programmer**

ESP32 board does not include micro-USB or USB connection as seen in Figure 2 so the FDTI is needed to supply power and code uploading. FDTI programmer can be seen in Figure 4. On the other hand, ESP32 and FDTI connection is shown in Figure 5.

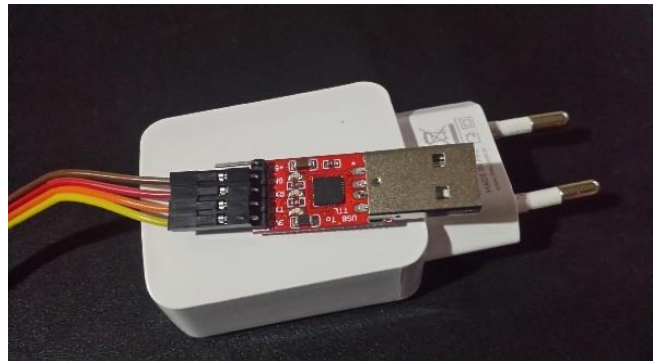

**Figure 4.** FDTI Programmer

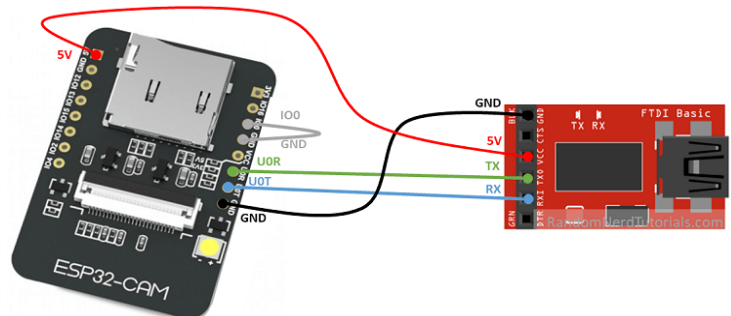

**Figure 5.** ESP32 and FDTI connection (Random Nerd Tutorials, 2021)

**ESP32 Camera Server**

ESP32-CAM comes with its own server. The image taken from the camera on the board is transferred to the webserver running on the board. In the webserver page, there is an interface where we can make settings on the received image. For example, resolution, brightness, contrast settings can be made as well as to reduce the size of the image data and the frame rate per second by using interface units. In Figure 6, camera server user interface is given.

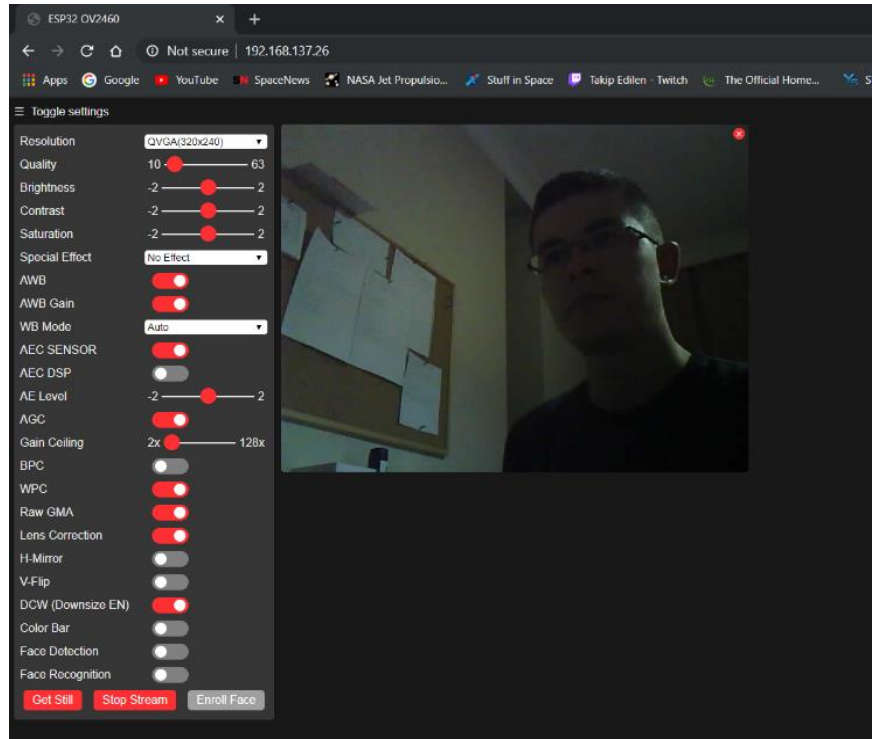

**Figure 6.** Camera Server UI

**PROPOSED WORK**

The server that comes in the ESP32 board basically works as follows. It forms the HTTP request that it wants to receive from ESP32 and transmits it to ESP32 via Wi-Fi network. When we press the "Start Stream" button from the server interface, this request is generated by running a JavaScript function. The function to be called is specified in the request. On ESP32, this function works and activates a flag. A thread that listens to this flag receives the buffer containing the image and sends it to the server through the "get" functions in the camera driver. The transmitted images are transmitted in "jpg" format and this transmission is performed without any encryption. The server takes the "jpg" file and reflects it on the screen.

The purpose here is to send the encrypted "jpg" file. However, as mentioned in the previous headings, the encryption algorithms to be executed should be small and fast due to the processor capacity and resource shortages of the device. Otherwise, the transmission will be delayed and the number of images per second will be reduced in the server-side video.

AES encryption algorithms have been found most suitable for this purpose and are also ideal for IoT devices. ECB (Electronic Codebook), CBC (Cipher Blocker Chaining) and CTR (Counter Mode) algorithms will be used for encryption because of their fast operation and low energy consumption.

There is an AES encryption library written in C called "Tiny AES". In this library, the encryption key length can be determined and applied by selecting three different encryption algorithms (ECB, CBC, CTR). The fact that it is written in C has increased both the efficiency and the resources of an IoT device. By default, a 128-bit key is defined, but a 192-bit and 256-bit key can also be specified. However, 128-bit key IoT is sufficient for video encryption on a device. 128 bit will be used as a result.

```c
/* Initialize context calling one of: */
void AES_init_ctx(struct AES_ctx* ctx, const uint8_t* key);
void AES_init_ctx_iv(struct AES_ctx* ctx, const uint8_t* key, const uint8_t* iv);

/* ... or reset IV at random point: */
void AES_ctx_set_iv(struct AES_ctx* ctx, const uint8_t* iv);

/* Then start encrypting and decrypting with the functions below: */
void AES_ECB_encrypt(const struct AES_ctx* ctx, uint8_t* buf);
void AES_ECB_decrypt(const struct AES_ctx* ctx, uint8_t* buf);

void AES_CBC_encrypt_buffer(struct AES_ctx* ctx, uint8_t* buf, uint32_t length);
void AES_CBC_decrypt_buffer(struct AES_ctx* ctx, uint8_t* buf, uint32_t length);

/* Same function for encrypting as for decrypting in CTR mode */
void AES_CTR_xcrypt_buffer(struct AES_ctx* ctx, uint8_t* buf, uint32_t length);
```

The capturePhoto() function located on server HTML file as a javascript function sends a request on the /capture URL to the ESP32, so it takes a new photo.

```
function capturePhoto() {
  var xhr = new XMLHttpRequest();
  xhr.open('GET', "/capture", true);
  xhr.send();
}
```

When the ESP32-CAM receives a request on the root **/** URL, ESP32-CAM send the HTML text to build the web page.

```
server.on("/", HTTP_GET, [](AsyncWebServerRequest * request) {
  request->send_P(200, "text/html", index_html);
});
```

When the capture button pressed (or start stream button) on the web server, a request sent to the ESP32 "/capture" URL. When that happens, we set the "takeNewPhoto" variable to "true", so that ESP32 knows it is time to take a new photo.

```
server.on("/capture", HTTP_GET, [](AsyncWebServerRequest * request) {
  takeNewPhoto = true;
  request->send_P(200, "text/plain", "Taking Photo");
});
```

After taking the photo, the AES library used to encrypt the image data and that encrypted data was used for the following processes. Using AES-128 for encryption, it is provided that secure transmission of the photo is successful.

**RESULTS AND DISCUSSION**

Internet of things devices appear to have security gaps due to their limited resources, their wide range of usage area and incomplete security standards. These devices, which are located in people's living areas, manufacturing and business processes also cause difficulties in protecting privacy. In this paper, we first mentioned briefly the security vulnerabilities of IoT devices and the fact that they do not have any prevention in terms of security. There is a lot of work to be done in the field of IoT security, and developers need to raise awareness to ensure security. For example, on a small IoT device such as the ESP32-CAM, we have indicated on the paper that this can be done. As a result, Tiny AES library was used with 128 encryption keys. Thus, a photo taken with ESP32 was encrypted and transmitted with AES, in the study so a reliable video broadcast system was proposed using ESP-Eye, an IoT device. The IoT device, which has a camera connected wirelessly to the network, broadcasts the recorded images with a server running on it.

In this study, it has been tried to provide secure video and image transfer by using ESP-Eye IoT device. Data is encrypted for security. AES, one of the symmetric encryption algorithms, is used in encryption operations. The study could also be carried out with a Raspberry-Pi, desktop computer or a high resolution camera that can be attached to any other computer system. Also, different encryption algorithms can be used instead of AES. However, it will be more effective to use the ESP-Eye device in many business operations because of its low cost, low energy consumption and high applicability. Especially for applications where the image quality of the ESP-Eye camera is sufficient, it stands out compared to its competitors due to the other advantages it offers. In addition, it is a great advantage that the AES algorithm can be applied with ESP-Eye. The AES algorithm is a security-strong algorithm. Although AES-128 is used, AES-256 usege can also be tried. With the proposed study, the security level of this technology, which can be used in many business processes such as logistics, quality control, production and planning, is increased.

In future studies, the performance of this study can be evaluated and the proposed study can be expanded with different IoT devices. In addition, its effectiveness can be measured by performing work tests in application areas.

**ACKNOWLEDGEMENT**

Thank you to Assoc. Prof. Dr. Gökhan DALKILIÇ and Dokuz Eylül University Computer Engineering Department for their support in the creation of this study.